\documentclass[journal]{IEEEtran}

\ifCLASSINFOpdf
 \else
\fi

\usepackage{wrapfig}
\usepackage{comment}
\usepackage{balance}  
\usepackage{cite}
\usepackage{graphicx}
\usepackage{float}
\usepackage{algorithm}
\usepackage{algpseudocode}
\usepackage{amsmath} 
\usepackage[utf8]{inputenc}
\usepackage[final]{pdfpages} 
\usepackage{pdfpages}
\usepackage{lipsum}
\usepackage{textcase}
\usepackage{url}
\usepackage{amsmath,esint}
\usepackage{epstopdf}
\usepackage{array} 
\usepackage{multirow}
\usepackage{booktabs}
\usepackage{esvect}
\usepackage{soul}
\usepackage[normalem]{ulem}
\usepackage{amsbsy}
\usepackage{amssymb}
\usepackage{subcaption}

\usepackage{textcomp}
\usepackage{multirow}
\usepackage{siunitx}
\DeclareSIUnit{\ppm}{ppm}

\newcolumntype{P}[1]{>{\centering\arraybackslash}p{#1}}
\newcolumntype{M}[1]{>{\centering\arraybackslash}m{#1}}

\begin{document}
%
% paper title

\title{Improvement of DVB-S2/S2X Performance Using External Synchronization}

\author{
\IEEEauthorblockN{Wahab Khawaja, Néstor J. Hernández Marcano, Rune Hylsberg Jacobsen\\
}
\IEEEauthorblockA{Department of Electrical and Computer Engineering, Aarhus University, 8000 Aarhus, Denmark}\\
Email: wahabgulzar@ece.au.dk, nestorj.hernandezm@gmail.com, rhj@ece.au.dk}

\maketitle

\begin{abstract}
Digital Video Broadcasting – Satellite, Second Generation (DVB-S2) and its extension DVB-S2X are widely used in modern satellite communications, where synchronization relies on physical layer headers, pilot symbols, and optional superframe structures but lacks defined implementation methods. This work explores the use of external synchronization to enhance DVB-S2 performance by using  GPS-disciplined oscillators (GPSDOs), and a hardware–software-in-the-loop satellite channel model emulating Low Earth Orbit (LEO) propagation. We evaluate scenarios with and without Doppler shifts and radio frequency~(RF) interference, comparing synchronized and unsynchronized cases. Results show that external synchronization significantly improves bit error rate~(BER), frame error rate~(FER), and signal-to-noise ratio~(SNR), subsequently reducing the frames required for reliable synchronization and enabling higher throughput in future satellite communication (SATCOM) systems.
\end{abstract}


\IEEEpeerreviewmaketitle

\section{Introduction}
The field of satellite communication~(SATCOM) has gained significant attention in recent years and is expected to play an even greater role in the future. Sixth-generation~(6G) and beyond networks are expected to integrate terrestrial and non-terrestrial systems, with SATCOM as a key component. According to \cite{goldman_sach}, the global satellite market is projected to grow nearly sevenfold, from the current $15$~billion to \$$108$ billion by $2035$. This rapid expansion is expected to deliver high-speed, low-latency, and ubiquitous internet and communication services worldwide. At the physical layer, Digital Video Broadcasting – Satellite, Second Generation (DVB-S2) and its extension DVB-S2X are among the most widely adopted standards, offering significant performance improvements over earlier systems~\cite{compare}. DVB-S2/S2X provide synchronization through physical layer headers, pilot symbols, and optional superframes~(for DVB-S2X only) and the implementation depends on the RX design. These fields allow the RX to lock frame timing, frequency, and sampling. The number of frames or symbols needed depends on factors such as signal-to-noise ratio~(SNR), Doppler, receiver (RX) design, and the selected profile. If headers or pilots are lost, bursts or time-sliced data may also be lost. In satellite systems, such losses are costly due to limited feedback, long delays, and short communication windows in Low Earth Orbit~(LEO), which reduce throughput per pass. Research on DVB-S2/S2X synchronization is limited, and this work addresses this gap by introducing and testing a novel synchronization setup under different scenarios.

\begin{figure*}[!t]
	\centering%\vspace{-2mm}
	\includegraphics[width=\textwidth]{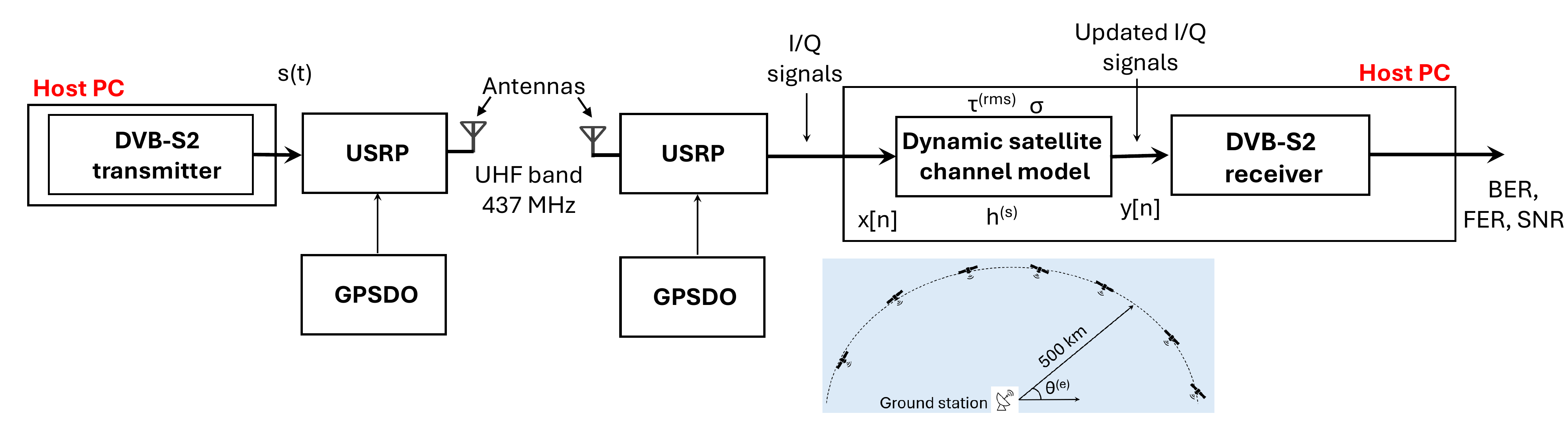}%\vspace{-3mm}
 \captionsetup{justification=raggedright, singlelinecheck=false}
	\caption{DVB-S2 transmission and reception setup through a LEO satellite channel using hardware and software in loop.}\label{Fig:Scenario}%\vspace{-5mm}
\end{figure*}

Several studies have addressed synchronization challenges in DVB-S2/S2X systems. In \cite{cite1}, frequency synchronization for LEO satellites was examined, showing that conventional geostationary orbit~(GEO) methods such as second-order phase-locked loop~(PLL) and Fitz fail under fast Doppler shifts, and proposing a low-complexity alternative. In \cite{cite2}, a frame synchronization method based on the scrambling sequence rather than the physical layer header was introduced, achieving better performance at low SNR due to the longer sequence length. A robust frame synchronization technique using differential generalized post-detection integration was presented in \cite{cite3}, offering resilience to large frequency offsets (up to 20\% of the symbol rate) while balancing complexity and performance. In \cite{cite4}, an improved coarse phase synchronization algorithm was proposed, tolerating nearly twice the residual frequency offset of previous methods and reducing FPGA complexity while maintaining bit error rate~(BER) below $10^{-7}$. Synchronization in DVB-RCS2 MF-TDMA burst systems was studied in \cite{cite5}, where oversampling with correlation and phase-difference with CRC validation enabled robust timing and carrier offset correction at low SNR. A phase-tracking block was designed in \cite{cite6} for DVB-S2X user terminals in precoding systems, optimizing a second-order PLL for phase-noise characteristics and validating its performance through Simulink modeling of superframe Format 2.

In this work, we developed a hardware- and software-in-the-loop satellite-to-ground propagation link using Universal Software Radio Peripherals~(USRPs) and a dynamic satellite channel model for the DVB-S2 communications. The testbed was evaluated using a typical DVB-S2 data-aided synchronization under scenarios with and without Doppler shift and with and without radio frequency~(RF) interference, comparing externally synchronized and unsynchronized~(internal USRP clock  only) cases. External synchronization was implemented using GPS-disciplined oscillators~(GPSDOs), as shown in Fig.~\ref{Fig:Scenario}. The results show significant improvements in BER, frame error rate~(FER), and SNR when external synchronization is applied in scenarios without Doppler shift, both with and without RF interference, while performance degrades under Doppler shifts. Overall, external synchronization demonstrates clear performance benefits for future DVB-S2/S2X-based satellite communications. To the best of our knowledge, this is the first work to integrate USRPs with external GPSDOs in a combined hardware- and software-in-the-loop DVB-S2/S2X testbed, providing a systematic comparison of synchronized and unsynchronized cases under realistic propagation conditions.

The rest of the paper is organized as follows: Section~\ref{Section:Methodology} provides the proposed methodology, the measurement setup is provided in Section~\ref{Section:Measurement}, the results and analysis is provided in Section~\ref{Section:Results}, and Section~\ref{Section:Conclusion} concludes the paper.

\section{Proposed Methodology} \label{Section:Methodology}
In this section, we present our proposed methodology for hardware and software-in-the-loop evaluation, data-aided synchronization for DVB-S2 (also applicable to DVB-S2X) and synchronization improvement using GPSDOs.

\subsection{Hardware-Software-in-Loop}
The novel architecture for satellite-to-ground DVB-S2 communications consists of two tightly coupled layers shown in Fig.~\ref{Fig:Scenario}. The implementation is shown for DVB-S2, however, can also be applied to DVB-S2X. The first layer is hardware-in-the-loop, implemented with USRP-based software defined radios~(SDRs) to transmit and 
receive DVB-S2 signals while capturing hardware impairments such as phase noise, frequency offsets, I/Q imbalance, antenna polarization mismatch, and RF front-end distortions (without the external power amplifier). The second layer is software-in-the-loop, which emulates the time-varying LEO satellite channel using statistical parameters derived from empirical measurements. This dynamic model accounts for large- and small-scale fading, Doppler shifts, and multipath components~(MPCs) that vary with satellite elevation angle during a pass, with the option to include weather-induced effects also. The channels in the two layers are considered linear.  

For an input signal $s(t)$, the hardware path through the SDR produces the baseband I/Q output
\begin{equation}
x^{(\rm h)}(t) = \big(h^{(\rm h)}(t) \circledast s(t)\big) + w^{(\rm h)}(t),
\end{equation}
where $h^{(\rm h)}(t)$ is the hardware CIR, dominated by a line-of-sight~(LOS) component with negligible MPC contribution compared to the LOS, $\circledast$ is the convolution operation, and $w^{(\rm h)}(t)$ is additive hardware noise. The sampled I/Q sequence at the time interval $T_{\rm s}$ is
\begin{equation}
x[n] \;=\; x^{(\rm h)}(nT_{\rm s}) \;=\; I[n] + j\,Q[n].
\end{equation}
This sequence is then passed to the software dynamic channel model prior to DVB-S2 decoding, 
as shown in Fig.~\ref{Fig:Scenario}.

\begin{figure*}[!t]
	\centering%\vspace{-2mm}
	\includegraphics[width=\textwidth]{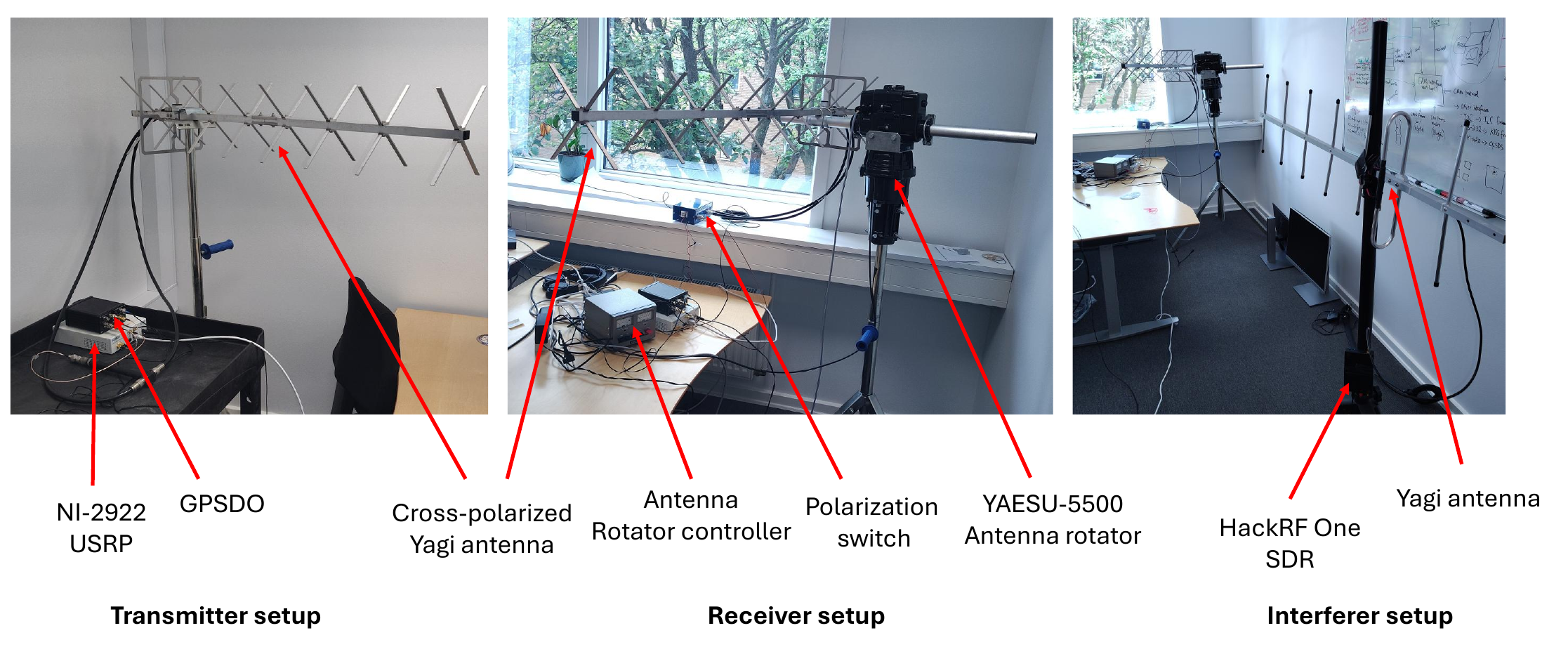}%\vspace{-3mm}
 \captionsetup{justification=raggedright, singlelinecheck=false}
	\caption{Indoor measurement setup with USRPs and GPSDOs; interferer with HackRF One and Yagi antennas.}\label{Fig:BER_15m}%\vspace{-5mm}
\end{figure*}

We have used 3GPP non-terrestrial network~(NTN) channel model for dynamic satellite channel modeling. 3GPP specifies different NTN profiles such as NTN-TDL-A, B, C, and D~\cite{3gppTR}. These profiles can represent satellite conditions at different elevation angles during a pass. In this work, we adopt the NTN-TDL-C profile, which includes both LOS and MPC components, to represent the suburban above-horizon channel~\cite{NTNTDLC} at a satellite elevation angle $\theta^{(e)}$. The satellite Doppler shift in the simulated channel model can be approximated as for a circular satellite orbit,  
\begin{equation}
f^{{\rm (D, sat)}} = \frac{v^{\text{(sat)}} f_{\rm 0}}{c} 
\cdot \frac{R_{\rm e} \sin\psi}{R_{\rm e} + h^{\text{(sat)}}} \cdot \rho,
\end{equation}
where $v^{(\rm sat)}$ is the orbital velocity of the satellite at altitude $h^{(\rm sat)}$, $f_{\rm 0}$ is the carrier frequency, $c$ is the speed of light, $R_{\rm e}$ is the Earth’s radius, $\psi$ is the central angle between the subsatellite point and the ground station, and $\rho \in \{+1,-1\}$ indicates whether the satellite is approaching or receding.
The continuous-time software channel output is
\begin{align}\label{eq:continuous}
y(t)
= &A_{0}\,x^{(\rm h)}(t-\tau_{0})\,e^{j\big(2\pi f^{(\rm D,0)}\,t + \phi \big)}
+ \nonumber \\ &\sum_{l=1}^{L-1}\big(h^{(\rm s)}_l \circledast x^{(\rm h)}\big)\!(t-\gamma_l) + w^{\mathrm{(h)}}(t),
\end{align}
where $A_0$, $f^{(\rm D,0)}$, and $\phi$ denote the amplitude, Doppler frequency shift, and initial phase of the LOS component, with $\phi \sim \mathcal{U}[0,2\pi)$. Here, $h^{(\rm s)}_l(t)$ is the time-varying CIR of the $l^{\rm th}$ path, $\gamma_l$ its relative delay, and $L$ is the total number of MPCs. Sampling at $t=nT_{\rm s}$ gives
\begin{align}\label{eq:integer_delays}
y[n] = &A_{0}\,x^{(\rm h)}[\,n-k_{0}\,]\,e^{j\big(2\pi f^{(\rm D,0)}\,nT_s + \phi\big)}
+\nonumber \\ &\sum_{l=1}^{L-1}\sum_{m=0}^{M_l-1} h^{(\rm s)}_l[m,n]\;x^{(\rm h)}[\,n - k_l - m\,] 
+ w^{\mathrm{(h)}}[n] ,   
\end{align}
where $k_l$ is the integer delay, $m$ the sample index within the $l^{\rm th}$ MPC, and $M_l$ the number of discrete time taps in path $l$.  If $
\mathbb{E}\!\big[\,|x^{(h)}[n]|^2\,\big] = 1$, signal power is given as
\begin{equation}
    P_{\text{sig}} \triangleq 
    |A_0|^2 \;+\; 
    \sum_{l=1}^{L-1} \sum_{m=0}^{M_l-1} 
    \mathbb{E}\!\left[ \, |h^{(\rm s)}_l[m,n]|^2 \, \right],
\end{equation}
then $\text{SNR} = \frac{P_{\text{sig}}}{P_{\rm w}}$, and $P_{\rm w} \triangleq \mathbb{E}\!\left[ |w^{(\rm h)}[n]|^2 \right] = \sigma_{\rm w}^2$ is the noise power.

At the RX, let $\mathcal{R}(\cdot)$ denote the physical-layer chain comprising DC offset removal, automatic gain control (AGC), symbol timing recovery, carrier frequency offset~(CFO) and phase compensation, frame synchronization, descrambling, constellation demapping, and forward error correction~(FEC) decoding using LDPC and BCH codes. The resulting baseband frames are then processed by $\mathcal{D}(\cdot)$, which performs stream and mode adaptation to reconstruct the transport stream in accordance with 
the DVB-S2 standard~\cite{DVBS2}. The recovered information bits $\hat{\mathbf{b}}$ and BER are given by
\begin{equation}
    \hat{\mathbf{b}} \;=\; \mathcal{D}\Big( \mathcal{R}\big( y[n] \big) \Big),
\quad \mathrm{BER} \;=\; \frac{1}{B}\sum_{i=1}^{B} \mathbf{1}\{\,b_i \neq \hat b_i\,\}, 
\end{equation}
where $B$ is the number of transmitted bits. FER is computed similarly. 
The effective throughput is 
\begin{equation}
R^{\rm (s)} \times n^{\rm(b)} \times r^{\rm (c)},    
\end{equation}
where $R^{\rm (s)}$ is the symbol rate, $n^{\rm(b)}$ the modulation order, and $r^{\rm (c)}$ 
the code rate.

\subsection{Data-Aided Synchronization for DVB-S2}
\label{sec:internal_sync}
Typical data-aided synchronization in DVB-S2/S2X is implemented in two stages. The first stage includes matched filtering, symbol timing acquisition \big(using a Gardner Timing Error Detector~(TED) and timing loop\big), frame synchronization through correlation with the physical layer header, and coarse carrier-frequency correction using either a frequency locked loop~(FLL) or a feed-forward estimator. These steps reduce frequency and timing offsets to a range where pilot-based fine estimators can operate effectively. The second stage then applies pilot-based frequency and phase estimation, interpolates phase estimates onto data symbols, and performs residual phase tracking to enable reliable decoding.

From (\ref{eq:integer_delays}), the baseband signal under frequency shift, sampling time error, and phase noise is given as 
\begin{align}
&y[n] = \nonumber \\ &e^{\,j \left( 2 \pi \Delta f \, n T_s + \varphi[n] \right)} \Bigg(
A_{0}\,x^{(\mathrm{h})}\Big[n - k_{0} - \frac{\tau[n]}{T_s}\Big]\, 
e^{\,j \left( 2 \pi f^{(\mathrm{D},0)} \, n T_s + \phi \right)} \;\; \nonumber \\
&+ \sum_{l=1}^{L-1} \sum_{m=0}^{M_l-1} h^{(\mathrm{s})}_l[m,n] \;
x^{(\mathrm{h})}\Big[n - k_l - m - \frac{\tau[n]}{T_s}\Big] 
\Bigg) + w^{(\mathrm{h})}[n],
\label{eq:yn_updated}
\end{align}
where \(\Delta f = f_{\mathrm{tx}} - f_{\mathrm{rx}}\neq 0\) is the CFO in Hz, $f_{\mathrm{tx}}$ and $f_{\mathrm{rx}}$ are the transmitter~(TX) and RX carrier center frequencies, \(\varphi[n]\) models phase noise from oscillator, \(\tau[n]\) is the timing offset from sampling clock offset~(SCO)-induced drift. If \(\epsilon_{\rm rx}\) and \(\epsilon_{\rm tx}\) represent the oscillator frequency offset in pulse per million~(ppm) at the RX and TX, respectively, we have
\begin{equation}
f_{\mathrm{rx}} = f_0(1+\epsilon_{\mathrm{rx}}), \qquad
\Delta f = f_0(\epsilon_{\mathrm{tx}}-\epsilon_{\mathrm{rx}}),
\end{equation} 
A generic FLL update used that helps in coarse carrier offset correction is given as
\begin{equation}
\widehat{f}_{k+1} = \widehat{f}_k + \beta_{\mathrm{FLL}}\,\varphi_k,
\label{eq:fll_revised}
\end{equation}
where $\widehat{f}$ is the frequency estimate, \(\varphi_k\) is a phase/frequency error indicator and \(\beta_{\mathrm{FLL}}\) is the loop gain. The coarse carrier frequency offset estimate~$CCFO$ is given as 
\begin{equation}
CCFO \approx \frac{\Delta \phi}{2\pi} \quad\Rightarrow\quad
\Delta f \approx CCFO\times R^{\text{(s)}},
\end{equation}
where \(\Delta\phi\) is the phase difference between samples spaced by one symbol.

If the RX sampling period is \(T_{\rm s}' = T_{\rm s}(1+\epsilon_{\rm s})\) with relative sampling error \(\epsilon_{\rm s}\), the timing error after \(n\) samples is approximately
\begin{equation}
\tau[n] \approx n T_{\rm s} \,\epsilon_{\rm s}.
\label{eq:sco_revised}
\end{equation}
The Gardner TED is used for timing error correction, where the scalar timing error estimate $\epsilon_k $ is given as
% Gardner TED in discrete-time (LaTeX)
\begin{equation}
\epsilon_k \;=\; \Re\!\left\{\, y\!\Big[n_k-\tfrac{N_{\rm s}}{2}\Big]\; \Big( y^*[n_k] - y^*[n_k-N_{\rm s}]\Big)\right\},
\label{eq:gardner_discrete_yn}
\end{equation}
where $\Re$ is the real operator, $N_{\rm s}$ is samples per symbol, and $n_k$ is sample index that corresponds to the decision instant for symbol $k$, $*$ is the complex conjugate operator, and the timing update follows a loop filter form given as
\begin{equation}
\tau_{k+1} = \tau_k + \beta_{\mathrm{tim}} \,\epsilon_k,
\label{eq:timing_update_revised}
\end{equation}
where \(\beta_{\mathrm{tim}}\) is the timing loop gain selected from the desired normalized loop bandwidth. Accurate interpolation is maintained when the signal is sampled at fractional delays.

%For frame synchronization, the start of frame is detected by correlating received symbols (after matched filtering and timing correction) with the known PL header sequence \(s_{\mathrm{header}}[\cdot]\) of length $N_{\rm header}$ given as
%\begin{equation}
%\Lambda(m) = \left|\sum_{n=0}^{N_{\mathrm{header}}-1} r[m+n]\; s_{\mathrm{header}}^*[n]\right|,
%\label{eq:frame_sync_revised}
%\end{equation}
%with energy normalization applied to make detection thresholds robust to SNR variation. 

For fine frequency and phase estimation using pilot blocks, phase estimates are computed per-pilot-block as
\begin{equation}
\widehat{\phi}_{\mathrm{pilot}}[k] = \arg\!\big( y_{\mathrm{pilot}}[k]\; p^*[k]\big),
\label{eq:pilot_phase_est_revised}
\end{equation}
where $y_{\mathrm{pilot}}[k]$ are the received pilot symbols, \(p[k]\) are the known pilot symbols, $*$ is the complex conjugate. A residual frequency estimate follows from the slope of pilot phases across known spacing \(N_{\rm p}\) given as
\begin{equation}
\widehat{f}_{\mathrm{pilot}} \approx \frac{\widehat{\Delta\phi}_{\rm pilot}}{2\pi\,N_{\rm p}\,T}.
\end{equation}
Pilot phase samples \(\widehat{\phi}_{\mathrm{pilot}}[k]\) are unwrapped and interpolated onto data symbol indices as follows
\begin{equation}
\widehat{\phi}_{\mathrm{data}}[n] = \mathrm{interpolate}\big(\widehat{\phi}_{\mathrm{pilot}}[k]\big),
\label{eq:phase_interp_revised}
\end{equation}
where linear interpolation is used.

While effective, the two stage synchronization approach has limitations. The major limitation is the dependence on the stability of the oscillators at the TX and RX sides for synchronization lock. Furthermore, the use of pilot symbols reduces spectral efficiency and adds overhead, at low code rates. Loop-based timing and carrier recovery may converge slowly and degrade under strong phase noise, or very low SNRs. Pilot-based interpolation can be insufficient for higher-order constellations or fast-varying channels, necessitating more complex data-aided loops. Moreover, the multi-stage frequency correction increases implementation complexity and may still leave residual offsets in some conditions.

\subsection{External Synchronization Using GPSDO}
\label{sec:gpsdo_sync_refined}
External GPSDOs are used at both the TX and RX to enhance synchronization for DVB-S2 transmission and reception. The data-based synchronization discussed in Section~\ref{sec:internal_sync} remains unchanged; however, with external GPSDOs disciplining both TX and RX references, the oscillator frequency errors $\epsilon_{\mathrm{tx}}, \epsilon_{\mathrm{rx}}$ are reduced from parts per million (ppm) to parts per billion (ppb) levels. As a result,
\begin{equation}
\Delta f \approx f_0(\epsilon_{\mathrm{tx}}-\epsilon_{\mathrm{rx}}) \;\;\to\;\; 0,
\end{equation}
and also and the sampling error $\epsilon_{\rm s}$ significantly reduces such that
\begin{equation}
\tau[n] \approx nT_{\rm s} \epsilon_{\rm s} \;\;\to\;\; \tau_0 \quad \text{(small constant offset)}.
\end{equation}
The received baseband signal in (\ref{eq:yn_updated}) simplifies to 
% Updated y[n] after internal sync (LaTeX)
\begin{align}
&y[n] = \nonumber \\ &e^{\,j\varphi[n]}\Bigg(
A_{0}\,x^{(\mathrm{h})}\Big[n - k_{0} - \frac{\tau_0 }{T_s}\Big]\,
e^{\,j\left(2\pi f^{(\mathrm{D},0)} n T_s + \phi\right)} + \nonumber\\
 &\sum_{l=1}^{L-1}\sum_{m=0}^{M_l-1} h^{(\mathrm{s})}_l[m,n]\;
x^{(\mathrm{h})}\Big[n - k_l - m - \frac{\tau_0 }{T_s}\Big]
\Bigg) +\; w^{(\mathrm{h})}[n],
\label{eq:yn_sync}
\end{align}
where $\varphi[n]$ is now a significantly small value and overall $y[n]$ shows the removal of large frequency ramps and timing drifts. A direct comparison is given in Table~\ref{Table:GPSDO}.

Using an external GPSDO at both the TX and RX substantially reduces CFO, SCO, and phase drift compared to internal synchronization, without adding data complexity. This reduction eases the load on DVB-S2/S2X synchronization loops, enabling narrower loop bandwidths, faster lock times, and greater robustness in low-SNR conditions. In this setup, the variance of CFO and timing estimates is driven primarily by thermal noise rather than oscillator instabilities. The lower drift also enhances the reliability of pilot-based phase estimation, which is important for high-order constellations.

\begin{table}[h]
\centering
\caption{Internal vs.\ GPSDO-based Synchronization}
\begin{tabular}{|c|c|c|}
\hline
\textbf{Variable} & \textbf{Internal Sync} & \textbf{With GPSDOs} \\
\hline
$\Delta f$ & $f_0(\epsilon_{\mathrm{tx}}-\epsilon_{\mathrm{rx}})$ & $\approx 0$ (Doppler only) \\
\hline
$\tau[n]$ & $nT_{\rm s} \epsilon_{\rm s}$ & $\tau_0$ (constant offset) \\
\hline
CCFO & Large search range & Narrow search range \\
\hline
Timing loop & Tracks linear drift & Tracks only small residuals \\
\hline
Phase $\phi[n]$ & Drift + noise & Short-term noise \\
\hline
\end{tabular}  \label{Table:GPSDO}
\end{table}

\section{Measurement Setup}  \label{Section:Measurement}
The measurements were conducted indoors using NI-2922 and NI-2920 USRPs with WBX-120 and SBX-120 daughterboards, as shown in Fig.~\ref{Fig:BER_15m}. Experiments were performed at $437$~MHz with two antenna configurations: a right-hand circularly polarized~(RHCP) cross-Yagi directional antenna and an ANT500 telescopic omnidirectional antenna with vertical polarization. At the TX, the two orthogonal elements of the cross-Yagi were fed with equal amplitudes and a $90^{\circ}$ phase shift, which was achieved by inserting a quarter-wavelength coaxial cable, to generate RHCP. At the RX, a polarization switch with equal-length external coaxial cables provided circular polarization from the two orthogonal antenna elements. The TX and RX were shown in Fig.~\ref{Fig:BER_15m}. A Fury GPSDO, which used a double-oven controlled crystal oscillator~(DOCXO) as its reference clock, was employed along with GPS antennas, with parameters listed in Table~\ref{Table:param}. Three scenarios were considered: (1) no Doppler shift and no RF interference, (2) residual Doppler shift without compensation, and (3) RF interference in the same band. Each scenario was tested for both synchronized (with external GPSDOs at the TX and RX) and unsynchronized cases.  

During the measurements, the sampling and symbol rate were kept fixed, and three fixed modulation and coding~(MC) schemes were used for two different transmit powers~(based on TX USRP gains) and two corresponding USRP RX gains. A fixed short FEC frame type of size $16200$ bits was used. A binary data matrix of size $1504\times \big(N_{\rm p}\times N_{\rm f}\big)$ was transmitted, where $1504$ was the number of bits per packet, $N_{\rm p}$ was the number of packets per frame for a given MC, and $N_{\rm f}$ was the number of frames per burst. The values of $N_{\rm p}$ were $4$, $6$, and $7$ for MC4, MC12, and MC24, respectively, and $N_{\rm f}=50$. For each MC scheme, ten iterations were performed under synchronized~(using external GPSDO) and unsynchronized~(internal USRP clock only) conditions. Each iteration consisted of transmitting $50$ frames as a burst. The coarse frequency estimator loop bandwidth $\beta_{\rm FLL}$ and the symbol timing synchronizer loop bandwidth $\beta_{\rm tim}$ are provided in Table~\ref{Table:param}, $\Delta f$, $\tau[n]$, $w[n]$, and pilot-based estimates are not explicitly assigned fixed values but are computed dynamically during runtime. Matlab was used for DVB-S2 signal generation, reception, and signal processing.  

\begin{table}[!h]
	\begin{center}
		\caption{Parameters for the hardware and software in loop.}\label{Table:param}
		\resizebox{\columnwidth}{!}{
        \begin{tabular}{@{} |P{6.2cm}|P{4.8cm}| @{}}
			\hline
			\textbf{Parameter} & \textbf{Parameter value} \\			
			\hline
		    Center frequency, $f_{0}$ & $437$~MHz \\
            \hline
            Antenna radiation pattern & Omnidirectional; directional (cross-Yagi) \\
            \hline
            Antenna gain & Omnidirectional: $2$~dBi; directional: $10$~dBi \\
            \hline
            Antenna polarization & Linear (vertical); RHCP \\
            \hline
            Transmit power & $16$~dBm \\
            \hline
            USRP RX gain & $30$~dB \\
            \hline
            Sampling rate, $F_{\rm s}$ & $2$~MSamples/s \\
            \hline
            Symbol rate, $R_{\rm s}$ & $1$~MSymbols/s \\
            \hline
            FEC frame type & short (16200 bits) \\
            \hline
            Carrier frequency estimator loop bandwidth, $\beta_{\mathrm{FLL}}$ & $0.8\times10^{-3}$  \\
            \hline
            Symbol timing synchronizer loop bandwidth, $\beta_{\mathrm{tim}}$& $0.6\times10^{-3}$  \\
            \hline
            MC: 4, 12, 24 & QPSK $\frac{1}{2}$; $8$PSK $\frac{3}{5}$; $32$APSK $\frac{3}{4}$ \\
            \hline
           Satellite altitude (orbital height) & $500$~km \\
            \hline
            Satellite elevation angle & $45^{\circ}$ \\
            \hline
            3GPP NTN channel profile & NTN-TDL-C (3GPP Rel.~17) \\
            \hline
            Shadowing (std.\ dev.), $\sigma_{\text{sh}}$ \cite{sd_lit}& $0.8$~dB \\
            \hline
            TDL (r.m.s.) delay spread $\tau^{\rm rms}$ \cite{dl_lit}& $80$~ns  \\
            \hline
            Uncompensated Doppler shift & $1$~kHz \\
            \hline
            RF interferer center frequency & $437$~MHz \\
            \hline
            RF interferer bandwidth & $300$~kHz \\
            \hline
            RF interferer transmit power & $0$~dBm \\
            \hline
            RF interferer antenna gain & $9$~dBi \\
            \hline
            RF interferer distance from RX & $2$~m \\
            \hline
            GPSDO fractional frequency stability & $1\times10^{-11}$ \\
            \hline
            GPS-locked timing accuracy & $\le 20$~ns \\
            \hline
        \end{tabular}
            }
	\end{center}
\end{table}

A dynamic LEO satellite channel model was adopted for a satellite orbiting at $500$~km, based on the 3GPP NTN-TDL-C channel profile, with a satellite elevation angle of $45^{\circ}$ above the horizon. The simulated channel uses empirical parameters: root mean square delay spread $\tau^{\rm (rms)} = 90$~ns, shadowing variance $\sigma = 0.8$~dB, and satellite velocity $v^{(\rm sat)} = 7.8$~km/s. The ground station was assumed to be located on a $23$~m high building in a suburban area. The channel model incorporated large-scale shadowing and small-scale fading following the 3GPP NTN-TDL-C profile. In the 3GPP NTN-TDL-C channel profile, the amplitude of the LOS component $A_0$ relative to the other MPCs is determined from the Rician $K$-factor defined in \cite{3gppTR}. While most of the Doppler shift was compensated at the RX, a residual Doppler component was retained for the Doppler shift scenario.  

An RF interferer with a Gaussian noise source directed toward the RX was implemented using a HackRF One SDR and a Yagi directional antenna. The interferer was configured with the parameters listed in Table~\ref{Table:param} and positioned as shown in Fig.~\ref{Fig:BER_15m}.

\section{Results and Analysis}   \label{Section:Results}
This section presents BER, FER, and SNR results for synchronized (using GPSDO clock and PPS sources) and unsynchronized (using USRP internal synchronization) scenarios. Measurements were carried out with omnidirectional and RHCP directional antennas across three MCs, two transmit power levels (TX USRP gains) and corresponding RX USRP gains. Results were reported for three scenarios: 1) without Doppler shift and interference called the clean scenario, 2) with residual Doppler shift scenario and 3) under RF interference . A BER or FER of $1$ indicated that all bits or frames were in error, while a BER of $10^{-8}$ represented the lower bound, corresponding to error-free decoding. This lower bound was used for empirical analysis since the number of bits and trials was finite and to avoid undefined condition.  

The normalized performance gain~(NPG) quantifies the relative improvement or degradation of BER and FER when comparing synchronized and unsynchronized scenarios. The NPG based on BER for comparing the synchronized and unsynchronized cases was defined as  
\begin{equation}
\text{NPG}_{\text{BER}} = \frac{\text{BER}_{\text{unsync}} - \text{BER}_{\text{sync}}}{\text{BER}_{\text{unsync}} + \text{BER}_{\text{sync}}}.
\label{eq:npg_ber}
\end{equation}
Similarly, the NPG for FER, $\text{NPG}_{\text{FER}}$, was obtained. The NPG metric ranged from $-1$ to $1$, where $1$ indicated maximum improvement (synchronization removed all errors), $-1$ indicated maximum degradation (synchronization introduced errors while unsynchronized reception was error-free), and $0$ indicated no net change.  

\begin{figure}[!t]
	\centering%\vspace{-2mm}
	\includegraphics[width=\columnwidth]{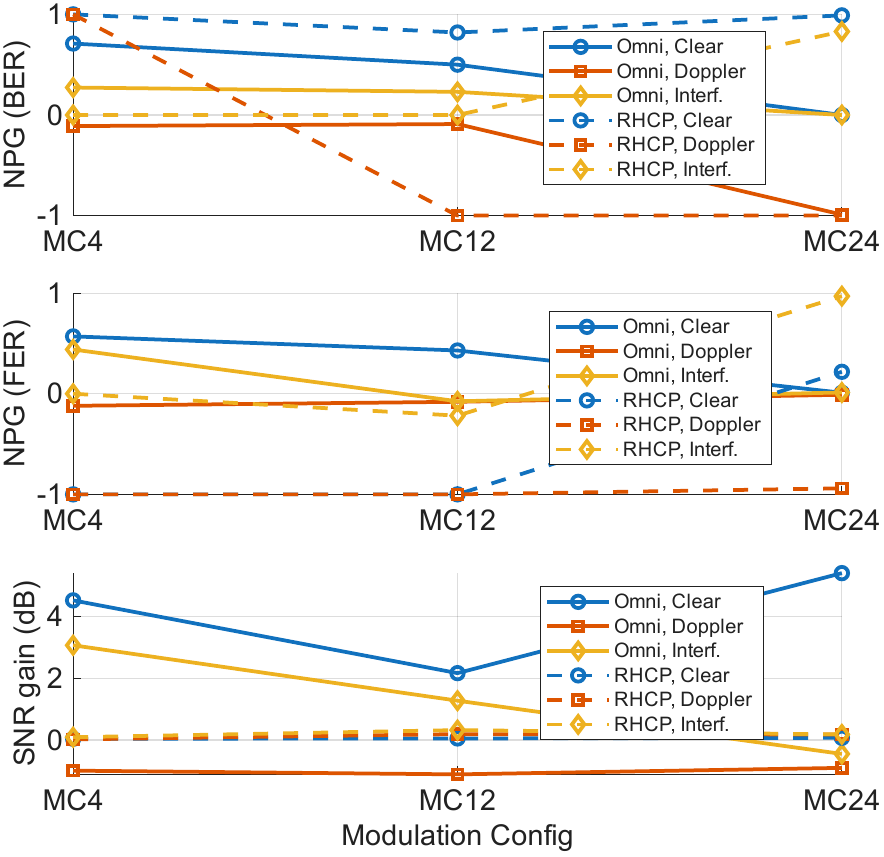}%\vspace{-3mm}
 \captionsetup{justification=raggedright, singlelinecheck=false}
	\caption{NPG (BER), NPG (FER), and average SNR gain for three scenarios and MCs for omnidriectional and RHCP directional antenna are shown.}\label{Fig:results_combined}%\vspace{-5mm}
\end{figure}

NPG based on BER and FER, along with the average SNR gain, was evaluated over $10$ iterations of 50-frame bursts for three MCs and three scenarios. Results for both omnidirectional and RHCP directional antennas under synchronized and unsynchronized cases are provided in Table~III.

\begin{table*}[h!]
    \centering
    \caption{Normalized performance gain (NPG) for BER and FER, along with average SNR gain for three MCs ($\mathrm{SNR}_{\mathrm{gain}} = \mathrm{SNR}_{\mathrm{sync}} - \mathrm{SNR}_{\mathrm{unsync}}$).}
    \label{tab:result-summary}
    \scalebox{1.0}{
    \begin{tabular}{|c|c|c|c|c|c|c|c|c|c|} \hline
         &  
      \multicolumn{3}{c|}{\textbf{NPG (BER)}}  &
      \multicolumn{3}{c|}{\textbf{NPG (FER)}}  &
      \multicolumn{3}{c|}{\textbf{SNR Gain (dB)}}  
      \\ \cline{2-10}
        & MC4 & MC12 & MC24 &
      MC4 & MC12 & MC24 &
      MC4 & MC12 & MC24 
      \\ \hline
      Omni, Clean & 
      0.71 & 0.50 & 0.00 &
      0.57 & 0.43 & 0.01 & 
      4.51 & 2.16 & 5.39 
      \\ \hline
      Omni, Doppler & 
      -0.11 & -0.09 & -0.99 &
      -0.12 & -0.08 & -0.01 & 
      -0.99 & -1.11 & -0.90 
      \\ \hline
      Omni, Interf. & 
      0.27 & 0.23 & 0.00 &
      0.44 & -0.08 & 0.01 & 
      3.06 & 1.27 & -0.45 
      \\ \hline
      RHCP, Clean & 
      1.00 & 0.82 & 0.99 &
      1.00 & 0.33 & 0.22 & 
      0.04 & 0.05 & 0.07 
      \\ \hline
      RHCP, Doppler & 
      -1.00 & -1.00 & -1.00 &
      -1.00 & -1.00 & -0.89 &
      0.02 & 0.19 & 0.18
      \\ \hline
      RHCP, Interf. & 
      0.00 & 0.00 & 0.83 &
      0.00 & -0.22 & 0.97 & 
      0.09 & 0.32 & 0.18 
      \\ \hline
    \end{tabular}
    }
\end{table*}

In addition, Fig.~\ref{Fig:results_combined} shows the BER- and FER-based NPG, along with the average SNR gain, comparing synchronized and unsynchronized cases.

Overall, the results show a positive NPG, indicating that synchronization improves performance in clean and under RF interference scenarios for both antenna types. For the omnidirectional antenna, the NPG decreases with higher MCs but remains positive, while for the RHCP antenna, the NPG increases with higher MCs, demonstrating that synchronization provides greater benefits at higher MC for the directional antenna shown in Fig~\ref{Fig:results_combined}.

In contrast, under residual Doppler shift, the NPG (both BER- and FER-based) is negative across all three MCs and antenna types, indicating that the synchronized case is more sensitive to Doppler compared to unsynchronized. Furthermore, majority of scenarios (except the Doppler shift) show positive SNR gain for the synchronized case showing the improvement achieved using the external synchronization. The SNR gain is larger for the omnidirectional antenna than the RHCP antenna. In the Doppler shift scenario with the omnidirectional antenna, the SNR gain is consistently negative, again indicating the sensitivity of the synchronized case under Doppler shift.

\section{Conclusions and Future Work}   \label{Section:Conclusion}
This work demonstrated that external synchronization in a DVB-S2 satellite-to-ground propagation link can significantly improve BER, FER, and SNR in scenarios without Doppler shift, both with and without RF interference. In contrast, performance degrades under uncompensated Doppler due to frequency locking effects. The gains are higher with omnidirectional antennas and vertical polarization than with directional RHCP antennas. Future work will investigate the quantized reduction in frame count resulting from external synchronization and assess its effect on throughput.

\ifCLASSOPTIONcaptionsoff
  \newpage
\fi

\balance
\bibliographystyle{IEEEtran}
\bibliography{References}

@INPROCEEDINGS{NTNTDLC,
  author={Zhu, Jingwen and Wang, Jie and Chen, Ming and Xue, Qingsheng and Tang, Xiangyuan},
  booktitle={Proc. IEEE Int. Conf. on Comput. and Commun. (ICCC)}, 
  title={{Time-domain Turbo Equalization for DVB-RCS2 Satellite Return Link in NTN-TDL Channel}}, 
  year={2023},
  volume={},
  number={},
  pages={956-961}}

@techreport{DVBS2,
  author       = {{ETSI}},
  title        = {Digital Video Broadcasting {(DVB); Second generation framing structure, channel coding and modulation systems for Broadcasting, Interactive Services, News Gathering and other broadband satellite applications (DVB-S2)}},
  institution  = {European Telecommunications Standards Institute (ETSI)},
  type         = {ETSI Standard},
  number       = {EN 302 307 V1.2.1},
  year         = {2009},
  month        = aug,
  address      = {Sophia Antipolis, France},
  url          = {https://www.etsi.org/deliver/etsi_en/302300_302399/302307/01.02.01_60/en_302307v010201p.pdf},
}

@ARTICLE{compare,
  author={Morello, A. and Mignone, V.},
  journal={Proc. of the IEEE}, 
  title={{DVB-S2: The Second Generation Standard for Satellite Broad-Band Services}}, 
  year={2006},
  volume={94},
  number={1},
  pages={210-227}}

@www{goldman_sach,
    author = {{Goldman Sachs}}, 
    title = {{The global satellite market is forecast to become seven times bigger}},
    url = {https://www.goldmansachs.com/insights/articles/the-global-satellite-market-is-forecast-to-become-seven-times-bigger},
    urldate = {30.08.2025},
}

@INPROCEEDINGS{cite1,
  author={Iansitov, Konstantin and Dorokhin, Semyon and Levichev, Sergey and Antiufrieva, Liubov and Dvorkovich, Alexander},
  booktitle={Proc. Int. Conf. Eng. and Telecommun. (En\&T)}, 
  title={{Low Complexity DVB-S2X Frequency Synchronization for LEO Satellites}}, 
  year={2021},
  volume={},
  number={},
  pages={1-5}
  }

@INPROCEEDINGS{cite2,
  author={Wu, Hao and Sha, Zhichao and Huang, Zhitao and Zhou, Yiyu},
  booktitle={Proc. IEEE Int. Symp. on Wireless Personal Multimedia Commun. (WPMC)}, 
  title={{Frame synchronization for DVB-S2 based on scrambling sequence}}, 
  year={2014},
  volume={},
  number={},
  pages={103-105}}

@INPROCEEDINGS{cite3,
  author={Kim, Pansoo and Corazza, G. E. and Pedone, R. and Villanti, M. and Chang, Dae-Ig and Oh, Deock-Gil},
  booktitle={Proc. IEEE Wireless Commun. and Netw. Conf.}, 
  title={Enhanced Frame Synchronization for {DVB-S2} System Under a Large of Frequency Offset}, 
  year={2007},
  volume={},
  number={},
  pages={1183-1187}}

@INPROCEEDINGS{cite4,
  author={Zhang, Xianyu and Zhang, Jing and Huang, Zixuan},
  booktitle={Proc. IEEE Int. Conf. on Intell. Comput. and Signal Process (ICSP)}, 
  title={An Improved {DVB-S2 Phase Coarse Synchronization Algorithm and Its FPGA Implementation}}, 
  year={2023},
  volume={},
  number={},
  pages={938-941}}

@INPROCEEDINGS{cite5,
  author={Xue, Qingsheng and Wang, Jie and Chen, Ming and Tang, Xiangyuan and Zhu, Jingwen},
  booktitle={Proc. IEEE Inf. Commun. Technol. Conf. (ICTC)}, 
  title={Efficient Receiver for {DVB-RCS2} Synchronization Problems}, 
  year={2023},
  volume={},
  number={},
  pages={167-171}}

@INPROCEEDINGS{cite6,
  author={Marrero, Liz Martínez and Duncan, Juan Carlos Merlano and Querol, Jorge and Chatzinotas, Symeon and Camps, Adriano and Ottersten, Björn},
  booktitle={Proc. IEEE Int. Conf. on Commun.}, 
  title={{A design strategy for phase synchronization in Precoding-enabled DVB-S2X user terminals}}, 
  year={2021},
  volume={},
  number={},
  pages={1-7}}

@techreport{3gppTR,
  author       = "{3rd Generation Partnership Project (3GPP)}",
  title        = "{{NR; User Equipment (UE) radio transmission and reception; Part 5: Satellite access (Release 17)}}",
  institution  = "3GPP / ETSI",
  number       = "TS 38.101-5, v17.11.0",
  year         = "2025",
  month        = "April",
  url          = "https://www.etsi.org/deliver/etsi_ts/138100_138199/13810105/17.11.00_60/ts_13810105v171100p.pdf"
}

@INPROCEEDINGS{dl_lit,
  author={Ning, Jiahao and Deng, Jinhao and Li, Yuanfang and Zhao, Chi and Liu, Jiashu and Yang, Songjiang and Wang, Yinghua and Huang, Jie and Wang, Cheng-Xiang},
  booktitle={Proc. IEEE Int. Conf. on Commun. in China (ICCC)}, 
  title={Ray-Tracing Channel Modeling for {LEO} Satellite-to-Ground Communication Systems}, 
  year={2024},
  volume={},
  number={},
  pages={1169-1174}}

@techreport{sd_lit,
  author       = "{{ITU Radiocommunication Sector (ITU-R)}}",
  title        = "{Propagation data required for the design of systems in the land mobile-satellite service}",
  institution  = "International Telecommunication Union (ITU)",
  type         = "Recommendation ITU-R P.681-11",
  number       = "P.681-11",
  series       = "P Series: Radiowave Propagation",
  year         = "2019",
  month        = "August",
  note         = "Available at: https://www.itu.int/rec/R-REC-P.681-11-201908-I/en"
}

%\appendix
%\section*{Appendix}
%\input{tables/appendix-tables}

\end{document}